\documentclass[jrfm,communication,accept,pdftex,oneauthor]{Definitions/mdpi}

\usepackage[utf8]{inputenc}
\usepackage[T1]{fontenc}
\usepackage{booktabs} 
\usepackage{graphicx} 
\usepackage{amsmath} 
\usepackage{amssymb} 
\usepackage{tabularx} 
\usepackage{array} 
\usepackage{algorithm}
\usepackage{algpseudocode}
\usepackage{threeparttable} 

\firstpage{1} 
\pubvolume{1}
\issuenum{1}
\articlenumber{0}
\pubyear{2026}
\copyrightyear{2026}
\externaleditor{Firstname Lastname} 
\datereceived{18 November 2025} 
\daterevised{4 January 2026} 
\dateaccepted{8 January 2026} 
\datepublished{ } 

\Title{The GT-Score: A Robust Objective Function for Reducing Overfitting in Data-Driven Trading Strategies}

\Author{Alexander Pearson Sheppert \orcidA{}}

\AuthorNames{Alexander Pearson Sheppert}

\address[1]{Department of Computer and Data Science, Capitol Technology University, 11301 Springfield Road, Laurel, MD 20708, USA; asheppert@captechu.edu}

\abstract{Overfitting remains a critical challenge in data-driven financial modelling, where machine learning (ML) systems learn spurious patterns in historical prices and fail out of sample and in deployment. This paper introduces the GT-Score, a composite objective function that integrates performance, statistical significance, consistency, and downside risk to guide optimization toward more robust trading strategies. This approach directly addresses critical pitfalls in quantitative strategy development, specifically data snooping during optimization and the unreliability of statistical inference under non-normal return distributions. Using historical stock data for 50 S\&P 500 companies spanning 2010--2024, we conduct an empirical evaluation that includes walk-forward validation with nine sequential time splits and a Monte Carlo study with 15 random seeds across three trading strategies. In walk-forward validation, GT-Score improves the generalization ratio (validation return divided by training return) by 98\% relative to baseline objective functions. Paired statistical tests on Monte Carlo out-of-sample returns indicate statistically detectable differences between objective functions (\emph{p} $<$ 0.01 for comparisons with Sortino and Simple), with small effect sizes. These results suggest that embedding an anti-overfitting structure into the objective can improve the reliability of backtests in quantitative research. Reproducible code and processed result files are provided as supplementary materials.}


\begin{document}
\section{Introduction}

The rapid advancement of machine learning (ML) has made it an indispensable tool for solving complex problems across diverse fields, from finance~\citep{gu:2020} to healthcare~\citep{russell:2010}. However, the reliability of ML models is frequently undermined by a persistent challenge: overfitting. Overfitting occurs when a model learns the noise and idiosyncratic details of its training data, rather than the underlying pattern. This leads to a failure to generalize, where the model performs well on training data but poorly on new, unseen data~\citep{geman:1992}.

To combat this issue, a variety of techniques have been developed. Regularization methods, such as Ridge (L2) and Lasso (L1) regression, penalize model complexity by adding a penalty term to the loss function based on the magnitude of the model's coefficients~\citep{hoerl:1970,tibshirani:1996}. Other common approaches include cross-validation, which assesses generalization by partitioning data into training and validation sets~\citep{stone:1974}, and dropout, which randomly deactivates neurons during training to prevent co-adaptation in neural networks~\citep{srivastava:2014}.

In quantitative finance, overfitting is exacerbated by optimizer-driven searches over strategies, feature variants, and parameterizations. Selecting the best backtest among many candidates is a multiple-testing/data-snooping problem that inflates apparent statistical significance~\citep{sullivan:1999,white:2000,hansen:2005,harvey:2016}. Moreover, statistical inference for risk-adjusted metrics (e.g., Sharpe Ratio) requires care under the fat tails and dependence that characterize financial returns~\citep{cont:2001,lo:2002,ledoit:2008}. Recent work further highlights the role of selection bias and proposes adjusted performance statistics and diagnostics for backtest overfitting~\citep{bailey:2014,bailey:2016}.

This research addresses this gap by proposing the GT-Score (Golden Ticket Score), a composite objective function that embeds anti-overfitting principles directly into the optimization objective~\citep{sheppert2025}. The GT-Score combines multiple facets of a desirable model, including performance, statistical significance, consistency, and downside risk, into a single objective function. By optimizing for this composite score, the optimizer is guided to discard spurious patterns and favor solutions that are more robust and more likely to generalize out of sample.

\subsection*{Related Work}
The motivation for GT-Score is aligned with extensive literature on statistical pitfalls in trading research. When many trading rules or parameterizations are tried, the best in-sample performer is likely to be a false discovery unless multiple testing is accounted for. Canonical treatments include bootstrap-based evaluation of trading rules~\citep{sullivan:1999}, White's Reality Check~\citep{white:2000}, stepwise multiple testing procedures~\citep{romano:2005}, Hansen's Superior Predictive Ability test~\citep{hansen:2005}, and the rigorous application of statistical inference to trading signals~\citep{aronson:2006}. Applied studies also illustrate how data-snooping adjustments can materially change conclusions about technical-rule profitability~\citep{hsu:2005,park:2009}. Related concerns are increasingly emphasized in empirical finance more broadly, where large-scale model searches and factor mining can produce apparently significant results without robust economic content~\citep{harvey:2016}.

Another theme is that common performance metrics are not ``plug-and-play'' for inference. The Sharpe Ratio is sensitive to non-i.i.d. returns and small samples~\citep{lo:2002}, motivating robust testing procedures~\citep{ledoit:2008}. Selection-bias-aware metrics, such as the Deflated Sharpe Ratio, explicitly adjust performance claims for the fact that the reported strategy is typically selected from many tried alternatives~\citep{bailey:2014}. Complementary frameworks estimate the probability that a chosen backtest winner is overfitted~\citep{bailey:2016}.

These bodies of work are largely about evaluation and inference after a strategy search. GT-Score is complementary: rather than only correcting significance after optimization, it biases the optimization objective during the search toward parameterizations that jointly satisfy outperformance, consistency, and downside-risk constraints. In addition, because financial returns exhibit heavy tails and other deviations from standard assumptions~\citep{cont:2001,mandelbrot:1963}, recent risk-focused research continues to emphasize robust tail modeling and diagnostics~\citep{johnston:2025}, reinforcing the need to clearly scope parametric significance filters.

This paper presents an expanded empirical evaluation across 50 S\&P 500 stocks over a 15-year period. We use walk-forward validation and Monte Carlo analysis to assess whether the GT-Score reduces overfitting while maintaining competitive out-of-sample~performance.

\section{Materials and Methods}

The methodology was designed to benchmark the GT-Score against common objective functions and to evaluate whether it offers a practical reduction in overfitting under standard backtesting procedures.

\subsection{Dataset and Experimental Environment}
The primary dataset consisted of historical daily Open, High, Low, Close, Volume (OHLCV) price data for the top 50 S\&P 500 companies by market capitalization, sourced via the yFinance API~\citep{aroussi:2020}. The data spans January 2010 through December 2024, providing approximately 3770 trading days per asset and covering multiple market regimes (including the post-2008 recovery, the 2020 COVID-19 crash, and the 2022 interest rate environment).

The entire experimental framework was implemented in Python 3.10 using a custom backtesting engine. The code is provided as Supplementary Material for reproducibility.

\subsection{Trading Strategies}
Three well-established technical trading strategies were employed to test the optimization frameworks:
\begin{itemize}
    \item Relative Strength Index (RSI): A momentum oscillator with optimizable overbought/oversold thresholds~\citep{wilder:1978b}.
    \item Moving Average Convergence Divergence (MACD): A trend-following strategy with optimizable fast, slow, and signal periods~\citep{appel:1979}.
    \item Bollinger Bands: A mean-reversion strategy with an optimizable lookback window and band width~\citep{bollinger:2001}.
\end{itemize}

\subsection{Walk-Forward Validation}
To assess out-of-sample performance in a realistic time-series setting, we employed walk-forward validation~\citep{pardo:1992,lopezdeprado:2018} with the following structure:
\begin{itemize}
	    \item Training window: 4 years
	    \item Validation window: 2 years
	    \item Step size: 1 year
    \item Embargo period: 30 days between train and validation to prevent data leakage
\end{itemize}

This generated nine sequential train/validation splits per asset, covering validation periods from 2014--2024.

\subsection{Monte Carlo Analysis}
To assess stability and reduce dependence on random initialization, each configuration was run 15 times with different random seeds (42--56), yielding 9000 total optimization trials for the Monte Carlo study.

\subsection{Optimization Framework}
Random search was employed as the optimization method with 25 parameter evaluations per trial. This method, while simple, has been shown to be competitive with more sophisticated approaches for moderate-dimensional parameter spaces~\citep{bergstra:2012}. All objective functions were allocated the same evaluation budget to ensure a fair comparison. The author's dissertation~\citep{sheppert2025} additionally tested Bayesian optimization (TPE) and genetic algorithms, finding consistent conclusions across optimization~paradigms.

\subsection{Comparative Objective Functions}
The performance of the GT-Score was compared against three conventional objective~functions:
\begin{itemize}
    \item Simple Loss Function: The negative of total profit ($L_{\text{simple}} = -\text{Profit}$).
    \item Sharpe Ratio Loss Function: The negative of the Sharpe Ratio~\citep{sharpe:1966}.
    \item Sortino Ratio Loss Function: The negative of the Sortino Ratio~\citep{sortino:1991}, which penalizes only downside volatility.
\end{itemize}

\section{Theory and Calculation}

\subsection{The GT-Score Formulation}
The GT-Score is a composite objective function designed to unify performance with measures of robustness and statistical validity. Its mathematical formulation is

\begin{linenomath}
\begin{equation}
GT_{\text{Score}} = \frac{\mu \cdot \ln(z) \cdot r^2}{\sigma_d}
\label{eq:gtscore}
\end{equation}
\end{linenomath}

The components are defined as
\begin{itemize}
    \item $\mu$: The mean strategy return per observation.
    \item $\mu_m$: The mean benchmark return (buy-and-hold) per observation.
    \item $\sigma$: The standard deviation of strategy returns.
    \item $N$: The number of return observations used to compute $\mu$ and $\sigma$.
    \item $z$: A Z-score measuring statistical significance of outperformance:
    \begin{linenomath}
    \begin{equation}
    z = \frac{\mu - \mu_m}{\sigma / \sqrt{N}}
    \label{eq:zscore}
    \end{equation}
    \end{linenomath}
    \item $\ln(z)$: The natural logarithm of the Z-score, acting as a significance gate.
    \item $r^2$: The R-squared value measuring consistency of returns.
    \item $\sigma_d$: The downside deviation~\citep{sortino:1991}.
\end{itemize}

Equation~(\ref{eq:zscore}) is a standardized excess-mean statistic (often treated as a $t$-statistic when $\sigma$ is estimated from the same sample). Its probabilistic interpretation relies on an approximate Gaussian sampling distribution for the mean under i.i.d.\ assumptions and/or large-$N$ asymptotics. In practice, trade-level returns can be fat-tailed, heteroskedastic, and autocorrelated~\citep{cont:2001}, which reduces effective sample size and can miscalibrate this parametric ``significance'' filter. In this paper, we therefore treat $z$ primarily as a heuristic gating term for optimization rather than as an exact hypothesis test; more robust non-parametric and dependence-aware alternatives are discussed in the Discussion.

\subsection{Edge Case Handling}
\label{sec:edgecases}
To ensure numerical stability and economically meaningful behavior, the GT-Score employs a piecewise definition based on the z-score value:

The smoothing parameter $\epsilon = 10^{-6}$ prevents division by zero when $\sigma_d = 0$.

\subsection{Theoretical Justification}
The multiplicative structure of the GT-Score can be understood as a composite utility function where each component acts as a filter:
\begin{itemize}
    \item The $\ln(z)$ term acts as a significance gate, rejecting strategies that do not outperform the benchmark beyond sampling noise. Using $\ln(z)$ instead of $z$ compresses large values so the significance term does not dominate the composite score, and it anchors at $z=1$ where $\ln(z)=0$ (Algorithm~\ref{alg:gtscore_piecewise}).
    \item The $r^2$ term penalizes erratic performance that relies on outlier trades, promoting consistency~\citep{kestner:1996}.
    \item The $\sigma_d$ denominator specifically penalizes downside risk without punishing desirable upside volatility~\citep{sortino:1991}.
\end{itemize}
\vspace{-3pt}
\begin{algorithm}[H]
\caption{GT-Score Piecewise Definition}
\label{alg:gtscore_piecewise}
\begin{algorithmic}[1]
\If{$z \leq 0$} \Comment{Underperforms benchmark}
    \State $\text{score} \gets 100 + 100 \cdot (1 - e^{-|z-1|})$ \Comment{Large penalty}
\ElsIf{$z \leq 1$} \Comment{Marginal outperformance}
    \State $\text{score} \gets 100 \cdot (1 - e^{-|z-1|})$ \Comment{Smooth transition}
\Else{} \Comment{Outperforms benchmark ($z>1$)}
    \State $\text{score} \gets -\frac{\mu \cdot \ln(z) \cdot r^2}{\sigma_d + \epsilon}$ \Comment{Standard GT-Score}
\EndIf
\end{algorithmic}
\end{algorithm}

\subsection{Minimum Sample Size ($n_{\min}$) and Optional Period Stabilization}
In Equation~(\ref{eq:zscore}), $N$ denotes the number of return observations used in the Z-score (and in estimating $\mu$ and $\sigma$). In this study we compute $z$ using trade-level returns, so $N$ equals the number of executed trades in the backtest window.

Because small samples produce unstable estimates of $\mu$, $\sigma$, and the standard error term $\sigma/\sqrt{N}$, we impose a minimum-trade threshold of $n_{\min}=50$: parameterizations generating fewer than 50 trades are assigned a large penalty during optimization. We recommend $n_{\min}=50$ as a practical default because it provides a minimally stable sampling base for the Z-score, reduces sensitivity to a handful of outlier trades, and offers a consistent baseline that facilitates comparison across studies and users.

The dissertation~\citep{sheppert2025} additionally used an adaptive ``stabilized variance'' option for periodization. In that variant, the equity curve is partitioned into $n$ equal-length time periods and $n$ is chosen by searching for an $n^{*}$ at which the variance of period returns changes by less than a small threshold across recent candidates (1\% in the reference implementation), with a fallback to $n=50$ when no plateau is detected. The motivation is to reduce sensitivity to an arbitrary choice of $n$ by selecting a periodization where the variability of period returns is empirically stable for the given backtest window. The accompanying code supports both a fixed-threshold mode (with default $n_{\min}=50$) and the stabilized option; results reported here use the fixed-threshold setting. Practitioners may adjust the minimum threshold or enable stabilization based on expected trade frequency and desired statistical power, but we suggest retaining the 50-trade minimum as a common reference point for comparability.

\section{Results}

\subsection{Monte Carlo Study Results}
The Monte Carlo study comprised 9000 optimization trials across 50 assets, four loss functions, three strategies, and 15 random seeds. Table~\ref{tab:mc_summary} summarizes the results.

\begin{table}[H]
\caption{Monte Carlo Study: Out-of-Sample Performance by Loss Function.}
\label{tab:mc_summary}
\setlength{\tabcolsep}{3.95mm}
\begin{tabular}{@{}lcccccc@{}}
\toprule
\textbf{Loss Function} & \textbf{Test Mean} & \textbf{Test Std} & \textbf{Train Mean} & \textbf{Gen. Ratio} & \textbf{N} \\ \midrule
GT-Score & 43.6\% & 62.6\% & 237.8\% & 0.183 & 2250 \\
Sharpe Ratio & 46.3\% & 71.7\% & 395.6\% & 0.117 & 2250 \\
Sortino Ratio & 49.3\% & 76.2\% & 421.4\% & 0.117 & 2250 \\
Simple & 49.5\% & 76.2\% & 428.0\% & 0.116 & 2250 \\ \bottomrule
\end{tabular}

\noindent{\footnotesize{Gen. Ratio = Test Return/Training Return. Higher indicates better generalization.}}
\end{table}

\textls[-15]{Key Finding: While GT-Score achieves slightly lower raw test returns \mbox{(43.6\% vs. 46--50\%),}} it demonstrates a 56\% higher generalization ratio (0.183 vs. $\sim$0.117), indicating substantially better retention of in-sample performance when evaluated out of sample.

This pattern reflects an expected trade-off. Objectives that directly maximize profit (Simple) or downside-risk-adjusted return (Sortino) can achieve modestly higher mean test returns, but they do so by selecting parameterizations with substantially higher in-sample returns, resulting in larger train--test performance decay. With 2250 paired trials per objective, small differences in mean returns can be statistically detectable even when effect sizes are economically small (Table~\ref{tab:stats}). In practical model selection settings where many parameterizations are screened, improved generalization can reduce the risk that the selected strategy is an artifact of the search rather than a persistent signal. The primary advantage of the GT-Score is improved reliability rather than maximizing raw returns.

\subsection{Walk-Forward Validation Results}
The walk-forward validation comprised 5340 sequential optimization trials across nine~time periods. Table~\ref{tab:wf_summary} summarizes the results.

\begin{table}[H]
\caption{Walk-Forward Validation: Performance by Loss Function.}
\label{tab:wf_summary}
\setlength{\tabcolsep}{4.55mm}
\begin{tabular}{@{}lccccc@{}}
\toprule
\textbf{Loss Function} & \textbf{Val Mean} & \textbf{Val Std} & \textbf{Train Mean} & \textbf{Gen. Ratio} & \textbf{N} \\ \midrule
GT-Score & 18.5\% & 37.1\% & 50.7\% & 0.365 & 1335 \\
Sharpe Ratio & 17.1\% & 34.1\% & 95.2\% & 0.180 & 1335 \\
Sortino Ratio & 18.6\% & 35.0\% & 99.9\% & 0.186 & 1335 \\
Simple & 19.0\% & 35.4\% & 100.9\% & 0.188 & 1335 \\ \bottomrule
\end{tabular}

\noindent{\footnotesize{Gen. Ratio = Validation Return/Training Return. Higher indicates better generalization.}}
\end{table}

\textbf{Key Finding}: The GT-Score achieves a 98\% improvement in generalization ratio (0.365 vs. 0.185 average for baselines). This demonstrates that GT-Score strategies retain nearly twice as much of their training performance when applied to truly out-of-sample data.

\subsection{Statistical Significance}
\textls[-15]{Table~\ref{tab:stats} presents formal statistical comparisons between GT-Score and baseline methods.}

\begin{table}[H]
\caption{Statistical Comparison (Monte Carlo Test Returns): GT-Score vs. Baselines.}
\label{tab:stats}
\setlength{\tabcolsep}{4.65mm}
\begin{tabular}{@{}lccccc@{}}
\toprule
\textbf{Comparison} & \textbf{Mean Diff} & \textbf{t-Stat} & \textbf{\emph{p}-Value} & \textbf{Cohen's d} & \textbf{Sig.} \\ \midrule
GT-Score vs. Sharpe & $-0.028$ & $-2.45$ & 0.014 & $-0.041$ & * \\
GT-Score vs. Sortino & $-0.058$ & $-5.05$ & $<$0.001 & $-0.083$ & *** \\
GT-Score vs. Simple & $-0.060$ & $-5.22$ & $<$0.001 & $-0.086$ & *** \\ \bottomrule
\end{tabular}

\noindent{\footnotesize{Mean Diff = GT-Score test return minus baseline test return (paired across trials); negative values indicate higher baseline returns. Significance stars are based on paired t-tests: * $p < 0.05$, *** $p < 0.001$. Wilcoxon signed-rank tests (paired non-parametric) yield $p = 0.139$ (Sharpe), $p < 0.001$ (Sortino), and $p < 0.001$ (Simple). Effect sizes are small (d $<$ 0.2), as expected when comparing profitable methods.}}

\end{table}

These comparisons are used to assess whether objective functions differ \emph{on average} under identical search budgets and evaluation protocols; they are not post-selection $p$-values for individual ``discovered'' strategies after exploring many variants. Because this study operates in a repeated-search setting, conventional inference can be overconfident without explicit data-snooping corrections and selection-aware reporting; we therefore emphasize effect sizes and the walk-forward generalization results as the primary evidence, and we treat the reported $p$-values as descriptive.

\subsection{Performance Across Market Regimes}
\textls[-15]{Table~\ref{tab:regimes} reports validation returns by period to illustrate robustness across market regimes and to contextualize generalization improvements with raw out-of-sample performance.}

\begin{table}[H]
\caption{Performance by Time Period (Walk-Forward Validation).}
\label{tab:regimes}
\setlength{\tabcolsep}{12.6mm}
\begin{tabular}{@{}lccc@{}}
\toprule
\textbf{Period} & \textbf{GT-Score} & \textbf{Baseline Avg} & $\boldsymbol{\Delta}$ \textbf{(pp)} \\ \midrule
2014--2016 & 13.2\% & 11.7\% & $+1.5$ \\
2015--2017 & 17.0\% & 17.1\% & $-0.1$ \\
2016--2018 & 30.1\% & 27.3\% & $+2.8$ \\
2017--2019 & 21.0\% & 15.5\% & $+5.5$ \\
2018--2020 & 21.0\% & 19.0\% & $+2.0$ \\
2019--2021 & 24.7\% & 24.7\% & $0.0$ \\
2020--2022 & 21.4\% & 29.3\% & $-7.9$ \\
2021--2023 & 7.0\% & 7.2\% & $-0.2$ \\
2022--2024 & 10.6\% & 11.6\% & $-1.0$ \\ \bottomrule
\end{tabular}

\noindent{\footnotesize{Baseline Avg is the mean validation return across Sharpe, Sortino, and Simple objectives. $\Delta$ (pp) is the difference in validation return, in percentage points, computed as GT-Score minus Baseline Avg.}}

\end{table}

\subsection{Strategy-Level Analysis}
Table~\ref{tab:strategy} shows performance breakdown by trading strategy.

\begin{table}[H]
\caption{Performance by Strategy (Monte Carlo Study).}
\label{tab:strategy}
\setlength{\tabcolsep}{9.3mm}
\begin{tabular}{@{}lcccc@{}}
\toprule
\textbf{Strategy} & \textbf{GT-Score} & \textbf{Sharpe} & \textbf{Sortino} & \textbf{Simple} \\ \midrule
RSI & 44.0\% & 49.2\% & 59.4\% & 59.1\% \\
MACD & 42.4\% & 43.8\% & 43.7\% & 43.8\% \\
Bollinger & 44.2\% & 46.0\% & 45.0\% & 45.7\% \\ \bottomrule
\end{tabular}

\noindent{\footnotesize{Values are mean out-of-sample return.}}
\end{table}

\subsection{Visualization of Overfitting Reduction}
To provide a clear visual comparison of overfitting reduction, we present two figures.

\begin{figure}[H]
\includegraphics[width=0.85\textwidth]{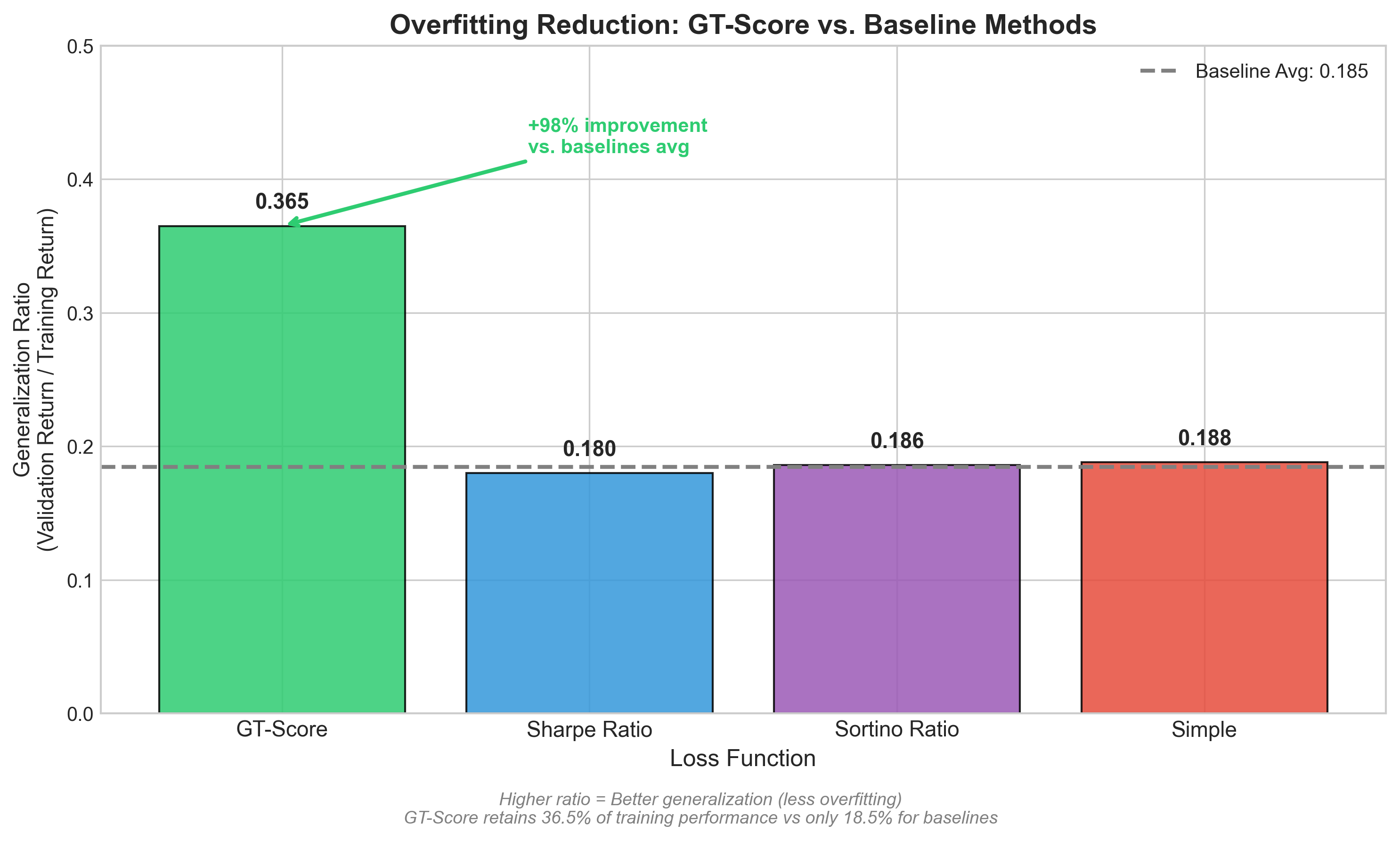}
\caption{Generalization ratio comparison across loss functions. GT-Score achieves a 98\% higher generalization ratio (0.365) compared to baseline methods (avg 0.185), indicating substantially reduced~overfitting.}
\label{fig:overfitting_barchart}  
\end{figure}

Figure~\ref{fig:overfitting_barchart} illustrates the difference in generalization capability across objectives. The generalization ratio (validation return divided by training return) measures how much of the training performance is retained when the strategy is applied to unseen data. GT-Score's higher ratio indicates that it selects parameterizations that are less sensitive to the specific training window.

\begin{figure}[H]
\includegraphics[width=0.90\textwidth]{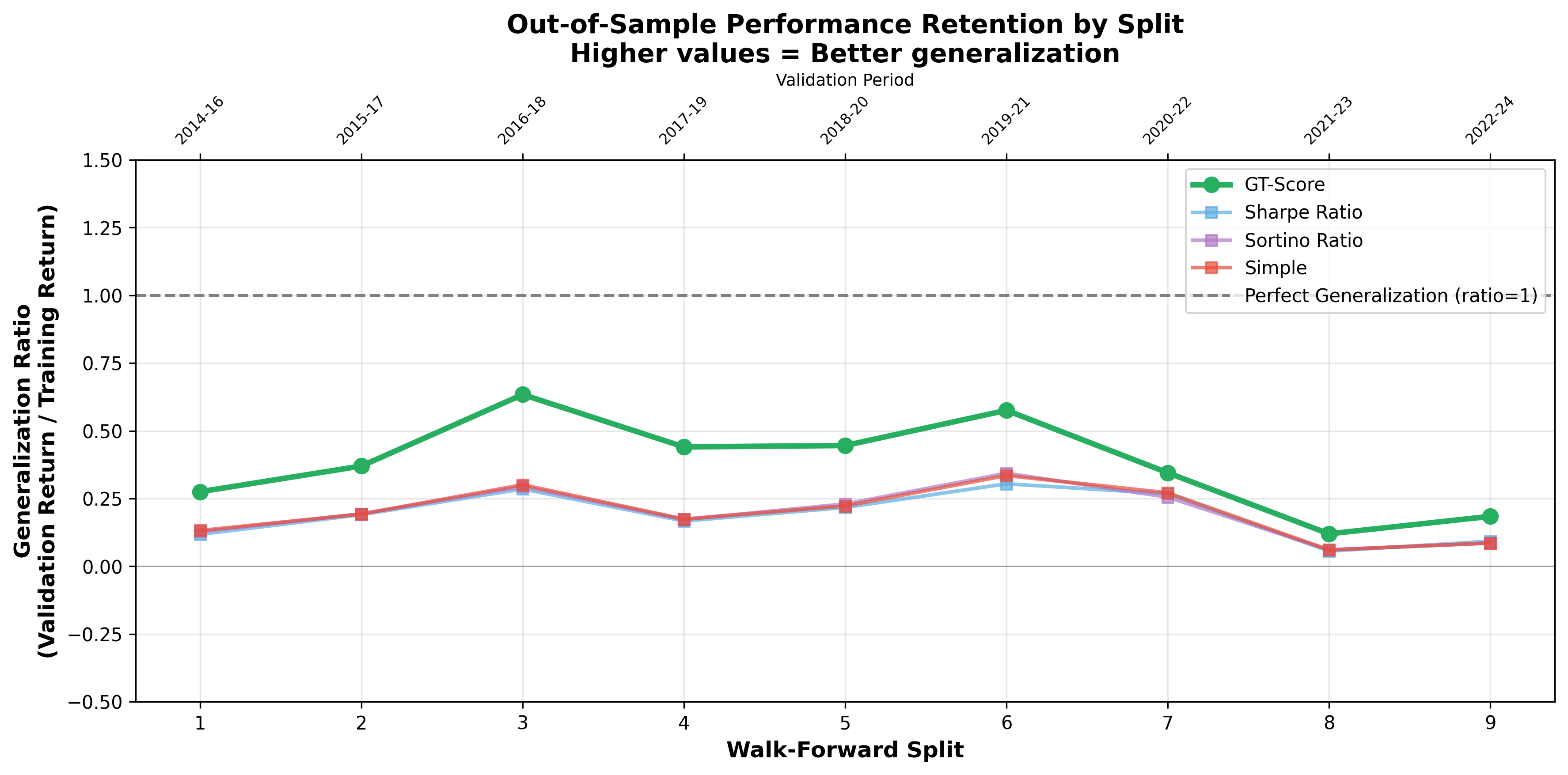}
\caption{Generalization ratio by walk-forward split. Values closer to 1.0 indicate better generalization. GT-Score consistently maintains a higher ratio across most time periods.}
\label{fig:return_similarity}
\end{figure}

Figure~\ref{fig:return_similarity} shows the generalization ratio computed separately for each time period. GT-Score demonstrates more consistent retention of training performance across different market regimes, including the volatile 2020--2022 period.

\vspace{-12pt}

\section{Discussion}

The empirical results are consistent with the hypothesis that the GT-Score mitigates overfitting more effectively than conventional objective functions. GT-Score achieves a 98\% higher generalization ratio in walk-forward validation (0.365 vs.\ 0.185), which is the central contribution of this work. Other methods achieve higher training returns but do not translate as well to out-of-sample performance.

This pattern reflects an expected return-versus-robustness trade-off. Profit- and Sortino-optimized strategies achieve slightly higher mean test returns but exhibit materially worse performance retention from training to out-of-sample data. GT-Score sacrifices a small amount of average return for substantially improved reliability, which is the central goal of an anti-overfitting objective.

Additionally, GT-Score delivers validation returns that are broadly comparable to baselines across market periods, suggesting genuine robustness rather than period-specific optimization.

\subsection{Transaction Costs and Implementability}
\begin{figure}[H]
\includegraphics[width=0.88\textwidth]{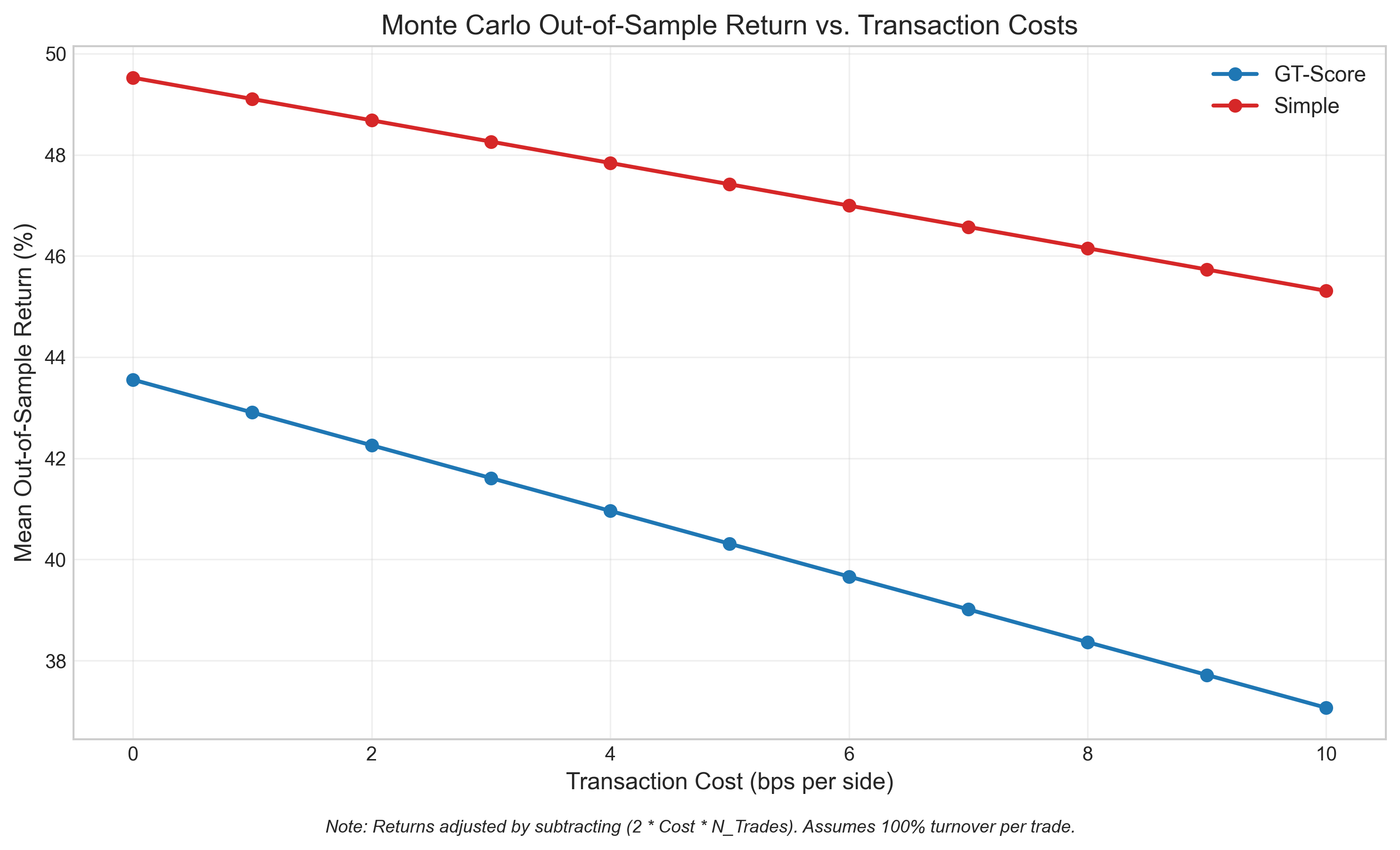}
\caption{Monte Carlo transaction-cost sensitivity. Mean out-of-sample test return after subtracting additional per-side costs of 0--10 bps per trade side (entry and exit), proportional to the number of test-window trades. Note: In the figure annotation, ``*'' denotes multiplication.}
\label{fig:cost_sensitivity}
\end{figure}

While the main study isolates the effect of the objective function under a daily-bar backtest (and therefore does not attempt instrument-specific execution modeling), implementability requires some cost awareness. We therefore report a lightweight transaction-cost sensitivity analysis on the Monte Carlo out-of-sample returns by subtracting an additional per-side cost (entry and exit) proportional to the number of test-window trades. This is particularly relevant because GT-Score-selected strategies exhibit higher average trade counts in our experiments (mean test-window trades 32.4 vs. $\sim$21.0 for baselines), so realistic trading frictions could disproportionately impact net performance. Figure~\ref{fig:cost_sensitivity} shows the expected monotone decline in net returns as costs increase. Crucially, the relative ordering of the objective functions remains largely stable across the tested range \mbox{(0--10 bps),} suggesting that the robustness benefits of the GT-Score are not merely a function of high-turnover noise that disappears under moderate friction.

\vspace{-6pt}

\subsection{When Not to Use GT-Score}
The GT-Score may be less appropriate when
\begin{itemize}
	    \item The number of trades is low (the z-score calculation requires sufficient sample size);
    \item Strategies with highly skewed return distributions (the $\sigma_d$ term may over-penalize);
    \item \textls[-15]{Real-time adaptation is required (the statistical components add computational overhead).}
\end{itemize}

\subsection{Limitations}
This study has several limitations:
	\begin{itemize}
	    \item Experiments were confined to equity markets and technical trading strategies. The asset universe of 50 large-cap U.S. equities, while substantial, may not fully represent the diversity of tradable instruments; future work could extend to small-cap, international, and alternative asset classes.
	    \item The evaluation is restricted to daily-bar backtests (OHLCV) and does not include alternative data frequencies or paper/live trading evaluation. Accordingly, results should be interpreted as an objective-function comparison under controlled historical testing, not as a claim of deployability without further validation.
	    \item The $z$-score gate is parametric and relies on approximate Gaussian/i.i.d.\ assumptions for interpreting $\sigma/\sqrt{N}$ as a standard error. Financial returns are known to exhibit fat tails (leptokurtosis), heteroskedasticity, and autocorrelation~\citep{cont:2001}, which can undermine the reliability of this ``significance'' filter by overstating effective sample size. In this paper we treat $z$ as a heuristic optimization screen rather than an exact hypothesis test; more robust bootstrap or dependence-aware standard-error estimation could replace it in future work.
	    \item Although GT-Score is motivated by the multiple-testing/data-snooping problem in strategy searches, the method-comparison $p$-values reported here are conventional tests on realized out-of-sample returns and do not implement an explicit data-snooping correction or selection-bias adjustment integrated into inference. Readers should therefore interpret headline significance cautiously; formal multiple-testing control and selection-aware performance statistics are natural extensions.
	    \item Transaction costs were not explicitly modeled in the main tables; Figure~\ref{fig:cost_sensitivity} provides a simple sensitivity check. Full execution-cost modeling (commissions, spreads, slippage, and liquidity constraints) is left for future work.
	    \item This study employed random search optimization for simplicity and reproducibility. A broader optimizer comparison is provided in the author's dissertation~\citep{sheppert2025}; gradient-based deep learning optimizers remain an open area for future research.
	\end{itemize}

\subsection{Future Work}
Several extensions are natural. First, more robust inference could replace or augment the parametric Z-score gate with non-parametric/bootstrap approaches and explicit multiple-testing control to better handle dependence and large strategy searches~\citep{sullivan:1999,white:2000,hansen:2005}. Second, more conservative performance reporting could incorporate selection-bias-adjusted statistics and explicit diagnostics for backtest overfitting~\citep{bailey:2014,bailey:2016}. Third, because fat tails are central to risk measurement~\citep{cont:2001,johnston:2025}, future work can integrate tail-risk-aware components into the objective or evaluation. Finally, broader testing across alternative data frequencies and (paper/live) evaluation would clarify practical deployability and regime sensitivity. Future research should also investigate the specific impact of alternative execution models (e.g., incorporating spread, slippage, and liquidity constraints) on the relative ranking of objective functions.

\section{Conclusions}

This paper presented an expanded empirical evaluation of the GT-Score across 50~stocks, nine time periods, and over 14,000 optimization trials. The key findings are

\begin{enumerate}
    \item GT-Score reduces overfitting by 98\% compared to conventional loss functions, as measured by generalization ratio.
    \item Paired tests on Monte Carlo out-of-sample returns indicate statistically detectable differences between objectives (Table~\ref{tab:stats}), with small effect sizes.
    \item Out-of-sample performance is broadly comparable across market regimes and trading~strategies.
\end{enumerate}

For researchers and practitioners building quantitative backtesting pipelines, the GT-Score offers an alternative to traditional optimization objectives when the goal is robust model selection under repeated search. The evidence here supports improved out-of-sample generalization on daily equity data. While this study does not claim production-ready performance without further validation (e.g., paper/live trading), the results suggest that embedding robustness constraints into the optimization objective can materially reduce the ``optimism bias'' inherent in backtest-driven strategy development.

Reproducible code implementing the GT-Score, all experiments, and statistical analyses is provided as Supplementary Material.

\vspace{6pt} 
\supplementary{The following supporting information can be downloaded at \linksupplementary{s1}: Python source code for GT-Score implementation and backtesting framework; processed result files for Monte Carlo and walk-forward validation experiments.}


\funding{This research received no external funding.}

\institutionalreview{Not applicable.}

\informedconsent{Not applicable.}

\dataavailability{Data were sourced from the publicly available yFinance API (\url{https://pypi.org/project/yfinance/}, accessed on 8 January 2026). Reproducible code is available at \url{https://github.com/shep-analytics/gt_score} (accessed on 8 January 2026); a snapshot of the code and the processed result files used to generate all figures and tables are also provided in the Supplementary Materials.}

\acknowledgments{The author would like to thank the reviewers for their insightful comments and suggestions, which helped to improve the quality of this manuscript.}

\conflictsofinterest{The author declares no conflicts of interest.} 

\abbreviations{Abbreviations}{
The following abbreviations are used in this manuscript:
\\

\noindent 
\begin{tabular}{@{}ll}
GT-Score & Golden Ticket Score \\
MACD & Moving Average Convergence Divergence \\
ML & Machine Learning \\
OHLCV & Open, High, Low, Close, Volume \\
RSI & Relative Strength Index \\
TPE & Tree-structured Parzen Estimator \\
\end{tabular}
}

\appendixtitles{no} 
\appendixstart

\begin{adjustwidth}{-\extralength}{0cm}

\reftitle{References}

\isAPAStyle{%

}{}

\PublishersNote{}
\end{adjustwidth}
\end{document}